\documentstyle[12pt,aasms4]{article}

\slugcomment{\shortstack[r]{Accepted by {\sl Monthly Notices of the Royal Astronomical Society}
}}

\begin{document}

\title
{How Galaxies Disguise Their Ages}

\author
{Roberto De Propris}

\affil
{School of Physics, Department of Astronomy \& Optics,
University of New South Wales, Sydney, Australia, NSW 2052; e-mail: 
propris@edwin.phys.unsw.edu.au}

\begin{abstract}

We calculate the contribution to Balmer line indices from far
ultraviolet component sources. We find that this is significant, and
may lead to identification of spurious age differences of the order of
a total span of $\sim 6$ Gyrs or $\sim 15\%$ size bursts observed a few
Gyrs after star formation stops. We suggest that claims for
intermediate age populations in early-type galaxies may need to be
reconsidered in the light of this new evidence.

\end{abstract}

\keywords{galaxies: evolution --- galaxies: stellar content}

\section
{Introduction}

For galaxies outside the Local Group, information on their stellar
populations can only be obtained from integrated properties, such as
colours and spectra. Unfortunately, integrated spectra and colours of
composite stellar populations are known to be degenerate with respect
to age and metallicity (e.g., Worthey 1994, for a recent discussion).
For instance, the existence of a tight color-magnitude correlation for
galaxies in nearby (Bower, Lucey \& Ellis 1992) and distant (Stanford,
Eisenhardt \& Dickinson 1998) clusters is taken to imply that cluster
galaxies form at high redshift over short timescales, whereas Schweizer
\& Seitzer (1992) are able to derive a tight $U-V$ vs. $V$ correlation
for field E/S0 galaxies, despite morphological and spectroscopic
evidence for the existence of young stellar populations in these
objects.

Worthey et al. (1994) have developed a system of narrowband spectrophotometric
indices useful for separating age and metallicity effects in integrated 
spectra. Line strengths of Balmer line indices (H$\beta$, H$\gamma$ and
H$\delta$) have been used to infer the presence of intermediate age
populations in some early-type galaxies (e.g., Worthey 1997 and references
therein). J{\o}rgensen (1999) uses H$\beta$ indices to infer the existence
of a correlation between mean age and metal abundance for galaxies in the
Coma cluster: young galaxies may masquerade among older objects because
of their higher metal abundance, allowing a wide range of ages to be
consistent with the passive evolution of the observed colour-magnitude 
relation and the small intrinsic scatter (Ferreras, Charlot \& Silk 1999).

On the other hand, the population synthesis models used to derive
theoretical index grids do not paint a complete picture of the stellar
populations of early-type galaxies. The spectral energy distributions
of these objects exhibit an unexpected rise in flux shortward of 2500
\AA, a phenomenon dubbed the far ultraviolet (FUV) upturn (see O'Connell
1999 for a review). The currently accepted explanation for the
FUV upturn is that it is caused by evolution of metal-rich stars on to
the extreme Horizontal Branch (HB) and their UV-bright progeny (Brown
et al. 1997 and references therein), whereas standard models terminate
their evolution on the red clump on the HB (e.g., Sweigart 1987). FUV
sources are known to exist in the metal rich, old ``open'' cluster
NGC6791 (Liebert, Saffer \& Green 1994) and are identified with field
subdwarfs O and B in our own Galaxy (Saffer et al. 1997). These objects
show strong, broad Balmer lines and may contribute significantly to
H$\beta$, H$\gamma$ and H$\delta$ indices. This is indeed the case for
some metal poor globular clusters with long blue tails (de Freitas
Pacheco \& Barbuy 1995).

Nevertheless, these stars are not included in the population synthesis
models of Worthey (1994), where HBs are treated as red clumps with a
temperature offset. The isochrones of Bertelli et al. (1994) used by
Vazdekis (1999) include an `AGB-manque' phase for high metallicities,
but only for low mass stars ($M < 0.60 M_\odot$) at large ages ($ > 20$
Gyr), whereas the existence of hot blue stars in NGC6791 and the field
sdO/B show that stars of $\sim 1 M_{\odot}$, are able to evolve on to
the extreme HB at ages of $\sim 10$ Gyrs (Carraro, Girardi \& Chiosi
1999). The models of Bressan, Chiosi \& Tantalo (1996) predict FUV
colours for galaxies of the appropriate ages and metallicities, but do
not calculate Balmer line indices contributed from FUV sources, since
fitting functions for stars of such high temperature and gravity are not
yet available. Worthey, Dorman \& Jones (1996) have computed the 
contribution to the integrated flux between 2000 and 2400 \AA\ from
a warm turnoff, and find that this may account for $\sim 50\%$ of the
observed luminosity, but do not explore the effect of HB sources on
spectrophotometric indices.

The purpose of this {\it Letter} is to consider, to a first
approximation, how FUV sources affect integrated Balmer line indices
and therefore whether the claims for younger mean ages in some
early-type galaxies may not be better explained by variations in HB
morphology. We find that these objects provide a significant amount of
Balmer line absorption and may therefore affect the derivation of ages via
the H$\beta$, H$\gamma$ and H$\delta$ indices.

\section{Modelling}

We adopt a semi-empirical approach in which we first estimate the fraction
of extreme HB stars (and progeny) needed to produce the observed FUV colours,
for two representative models at very different metallicities, and then
compute the contribution from these objects to the total Balmer line
absorption strength.

We use models by Dorman, O'Connell \& Rood (1993a) in which a 10--30\%
fraction of HB stars evolves on to UV-bright phases. We use the two models
for which detailed evolutionary calculations are presented by Dorman,
Rood \& O'Connell (1993b): one with [Fe/H]=+0.38, Y=0.292, $M_c=0.464
M_{\odot}$ (where $M_c$ is the core mass) and envelope masses of 0.003,
0.046 and 0.096 $M_{\odot}$ and a model with [Fe/H]=--1.48, Y=0.247,
$M_c=0.485 M_{\odot}$ and envelope masses of 0.003, 0.035 and 0.105
$M_{\odot}$. For each of these models we follow the prescriptions of
Dorman et al. (1993a) to calculate FUV colours and their contribution
to the total $V$ band light.

We then use the evolutionary tracks presented in Dorman et al. (1993b)
to estimate the fraction of total light at each $T_{eff}$ and $\log g$
step and the stellar atmospheres of Kurucz (1993) to estimate the
equivalent widths of H$\beta$, H$\gamma$ and H$\delta$ at each stage.
Since most of the models reach temperatures and gravities in excess of
the grid calculated by Kurucz, we extrapolate H$\beta$,$\gamma$,
$\delta$ equivalent widths to the appropriate T$_{eff} - \log g$ range
by means of polynomial fits to the predictions for the existing grid,
being unable to completely simulate the spectrum of these objects.
These are then summed, scaling by the fraction of total light produced,
to yield the equivalent widths of Balmer lines contributed from FUV
sources during their lifetime and again scaled by the fraction of total
$V$ band light to calculate the additional absorption line strength to
integrated H$\beta$, H$\gamma$ and H$\delta$ indices, following the
prescriptions of Freitas Pacheco \& Barbuy (1995). Table 1 shows the
models used. Here column 1 is the fraction of blue HB stars, column 2
the contribution to the $V$ band light, column 3 the 1550$-V$ color,
column 4 the 2500$-V$ color and column 5 the extra H$\beta$ strength
provided by these stars.  A header at the top indicates the model
parameters and mean temperature of the models.

\section{Discussion}

Figure 1 shows the excess equivalent width of H$\beta$ produced by HB
stars and their progeny as a function of FUV colour of the host galaxy.
We only plot the extra contribution to Balmer line strengths provided
by these stars, without assuming any underlying model. We assume that
the Balmer line strength produced by FUV sources can be added linearly
to the chosen galaxy model from the Worthey (1994) compilation.

The range of FUV colours in the Burstein et al. (1988) sample is 2 to
4.  From Figure 1, this corresponds to H$\beta$ equivalent widths of up
to 0.6 \AA, with a typical contribution of 0.3 \AA, in agreement with
earlier results on metal poor globular clusters (de Freitas Pacheco \&
Barbuy 1995) and `blue HB' simulations of Buzzoni, Mantegazza \&
Gariboldi (1994), albeit at lower temperatures than those of FUV
sources.  In our simulations, metal-poor HB stars yield somewhat higher
H$\beta$ strengths than those of metal-rich stars.

Figure 2 plots a single stellar population (SSP) model grid from
Worthey (1994), and superposes the range of H$\beta$ strengths
contributed by FUV sources to an underlying 12 Gyr old population.  We
add index strengths for the SSP and the HB stars linearly, as stated
above. We also show some of the higher signal-to-noise measurements of
Trager et al. (1998). It can be seen that spurious age differences of
$\sim 5-7$ Gyrs can be introduced by FUV sources; conversely, assuming
that more complex stellar populations can be represented by linear
combinations of SSP models, a 10--20\% burst of star formation observed
$1-3$ Gyrs after star formation ceases may be explained by FUV
sources.  Larger bursts observed at later ages can also be accounted
for in this manner.

Figure 1 shows that a range of H$\beta$ strengths is possible at any
FUV colour. In turn, the spread in FUV colours at any age larger than
5 Gyr is significant (Tantalo et al. 1996). Taken at face value, this
suggests some degeneracy in FUV colour, H$\beta$ strength and age for
early-type galaxies.

Our simulations show that the contribution to H$\gamma$ and H$\delta$
are similar to H$\beta$. The model grid of Jones \& Worthey
(1995) spans about 0.2 \AA, for a range of ages from 3 to 17 Gyrs.
This suggests that FUV sources strongly affect these indices as well.
On the other hand, the narrow H$\gamma_{A,F}$ and H$\delta_{A,F}$
indices of Worthey \& Ottaviani (1997) appear to be much less sensitive
to FUV source contributions, although in this case the narrowness of 
the index bandpasses probably requires more accurate modelling. The
new H$\gamma$ index of Vazdekis \& Arimoto (1999) varies by about 0.5
\AA\ over ages of 1.6 to 17 Gyrs (depending on velocity dispersion).
After correction for velocity dispersion, FUV source contribution can
vary between 0 and 0.4 \AA, spanning a sizeable portion of these
newer grids.  

Since Dorman et al. (1993b) do not present integrated fluxes for bands
other then $V$, we are unable to estimate the effect of FUV sources on
broadband colours.  Nevertheless, the contribution to $V$ from high
metallicity models is typically 5\% and always less than 10\%, which
should not affect broadband colors. For low metallicity models HB stars
may provide as much as 20\% of the light and this may produce bluer
than expected colours. For comparison, the two globular clusters with
long blue tails, M13 and NGC6229, are seen to have too blue integrated
$U-B$ for their $B-V$ colour (Reed, Hesser \& Shawl 1988).

An useful consistency check is to compare $2500-V$ colours for our
models and observations. For high metallicity systems, Table 1 shows
that we reproduce well the observed range of colours in the sample of
Burstein et al. (1988), which is typically 3--4. Low metallicity models
are far bluer, which is not surprising, since early-type galaxies are
generally metal rich, but they are consistent with $2500-V$ colours of
globular clusters, which are typically 1.5--3.5 and span the
appropriate metallicity range.

Figure 3 plots excess H$\beta$ emission (over an 18 Gyr old population)
for galaxies with index measurements from Trager et al. (1998) and FUV
colours from Burstein et al. (1988). This was calculated following
Davies, Sadler \& Peletier (1993) and includes some galaxies for which
they provide excess H$\beta$ values. We find no significant correlation
between $\Delta$H$\beta$ and 1550$-V$ colour, although some galaxies
have both strong excess absorption and blue FUV colors.  One of these
galaxies, however, is NGC5102, the well known E+A galaxy. 

This may imply that the effect of FUV sources on derived index
strengths is weak. As shown in Table 1, the largest contribution to
H$\beta$ comes from objects with high surface temperature. Brown et al.
(1997) estimate that the largest contribution to the FUV flux comes
from stars with surface temperatures lower than 25,000 K. These sources
would not provide large H$\beta$ absorption, in agreement with the data
presented in Figure 3. One caveat, is that FUV colors are measured
through the large IUE aperture, which is generally wider than the slit
sizes used to measure spectral indices: Ohl et al. (1998) show that FUV
color gradients can be strong, making a comparison such as shown in Figure 3
possibly unrealistic.

We have chosen two of the Dorman et al. (1993b) tracks considered to
best represent the FUV sources: it should be noted, however, that 
different objects, or different evolutionary tracks, may contribute
to the FUV upturn: for instance post AGB stars are believed to be
important in M31 (Bertola et al. 1995). The range of models used needs
to be considerably expanded to include a wider range of objects to
make these results more comprehensive.

There are a number of observational tests of our models: it is possible
to measure indices for the `quiescent' samples of Burstein et al.
(1988) and Longo et al. (1989) in consistent apertures. Conversely, FUV
strengths can be derived, from HST data, for Coma galaxies observed by
J{\o}rgensen (1999). It is also possible to explore the correlation of
line strength indices with the distribution of FUV components observed
by UIT (Brown et al. 1997). More accurate stellar atmospheres and
fitting functions for hot, high gravity, metal rich stars are also
necessary, to allow the FUV component effects to be accounted for in
population synthesis models.

\section{Conclusions}

We have considered the FUV source contribution to Balmer line indices
using a semi-empirical modelling technique. We find that this is significant
and may lead to identification of spurious intermediate age populations or
age gradients in galaxies. Nearly all Balmer line indices in use are 
potentially affected by FUV stars and consideration should be paid
to their age sensitivity.

\acknowledgments

We would like to thank Adam Stanford for commenting on an earlier version
of this paper, and Sabine M{\"o}hler for her help with Kurucz models.
We would also like to thank the referee, Dr. Guy Worthey, for a number
of useful suggestions. This work is supported by the Australian Research
Council.

\clearpage
\centerline {FIGURE CAPTIONS}

Figure 1: FUV colours vs H$\beta$ strength produced by our models.
The shaded region represents the range of FUV colours allowed by 
our simulations. A range of Balmer line strengths is possible at
each colour. 

Figure 2: A single stellar population grid of H$\beta$ vs. [Fe/H]
for different ages and abundances. We superpose the range of possible contributions
from FUV sources to a 12 Gyr old population (dark bars). We also plot
indices for high goodness galaxies (open circles) from the sample of Trager et al.
(1998). Age differences of a few Gyrs, for old populations, or bursts
of 10--20\% size observed 1--3 Gyrs after star formation ceases, can be
accounted for by FUV sources.

Figure 3: Excess H$\beta$ emission vs. FUV colour, from the sample
of galaxies in common to Trager et al. (1998), Davies et al. (1993)
and Burstein et al. (1988). Excess H$\beta$ is defined as in Davies et
al. (see text).

\clearpage 

\begin{deluxetable}{cccccccccc}
\tablecaption{The Models}
\tablehead{
\colhead{[Fe/H]} & \colhead{Y} & \colhead{M$_c$} & \colhead{M$_{env}$} &
\colhead{$<T_{eff}>$} & \colhead($f_b$) & \colhead{$V$ contr.} & \colhead{1550$-V$} &
\colhead{2500$-V$} & \colhead{H$\beta$ (\AA)}}
\startdata
0.38 & 0.29 & 0.464 & 0.003 & 51000 & 0.1 & 0.02 & 2.57 & 4.10 & 0.07 \nl
0.38 & 0.29 & 0.464 & 0.003 & 51000 & 0.2 & 0.04 & 2.02 & 3.75 & 0.14 \nl
0.38 & 0.29 & 0.464 & 0.003 & 51000 & 0.3 & 0.05 & 1.64 & 3.48 & 0.21 \nl 
0.38 & 0.29 & 0.464 & 0.046 & 28000 & 0.1 & 0.03 & 3.15 & 3.99 & 0.13 \nl
0.38 & 0.29 & 0.464 & 0.046 & 28000 & 0.2 & 0.06 & 2.80 & 3.61 & 0.26 \nl
0.38 & 0.29 & 0.464 & 0.046 & 28000 & 0.3 & 0.10 & 2.54 & 3.32 & 0.39 \nl 
0.38 & 0.29 & 0.464 & 0.096 &  3000 & 0.1 & 0.02 & 3.62 & 4.66 & 0.01 \nl
0.38 & 0.29 & 0.464 & 0.096 &  3000 & 0.2 & 0.05 & 2.88 & 4.75 & 0.02 \nl
0.38 & 0.29 & 0.464 & 0.096 &  3000 & 0.3 & 0.07 & 2.43 & 4.84 & 0.03 \nl 
--1.48 & 0.25 & 0.485 & 0.003 & 56500 & 0.1 & 0.06 & 2.98 & 2.07 & 0.17 \nl
--1.48 & 0.25 & 0.485 & 0.003 & 56500 & 0.2 & 0.12 & 2.32 & 2.01 & 0.34 \nl
--1.48 & 0.25 & 0.485 & 0.003 & 56500 & 0.3 & 0.18 & 1.93 & 1.94 & 0.51 \nl 
--1.48 & 0.25 & 0.485 & 0.035 & 23000 & 0.1 & 0.06 & 2.91 & 2.05 & 0.18 \nl
--1.48 & 0.25 & 0.485 & 0.035 & 23000 & 0.2 & 0.12 & 2.37 & 1.96 & 0.36 \nl
--1.48 & 0.25 & 0.485 & 0.035 & 23000 & 0.3 & 0.17 & 2.00 & 1.87 & 0.54 \nl
--1.48 & 0.25 & 0.485 & 0.105 &  4000 & 0.1 & 0.02 & 3.35 & 2.01 & 0.02 \nl
--1.48 & 0.25 & 0.485 & 0.105 &  4000 & 0.2 & 0.05 & 2.95 & 1.91 & 0.04 \nl
--1.48 & 0.25 & 0.485 & 0.105 &  4000 & 0.3 & 0.07 & 2.65 & 1.80 & 0.06 \nl
\enddata
\end{deluxetable}

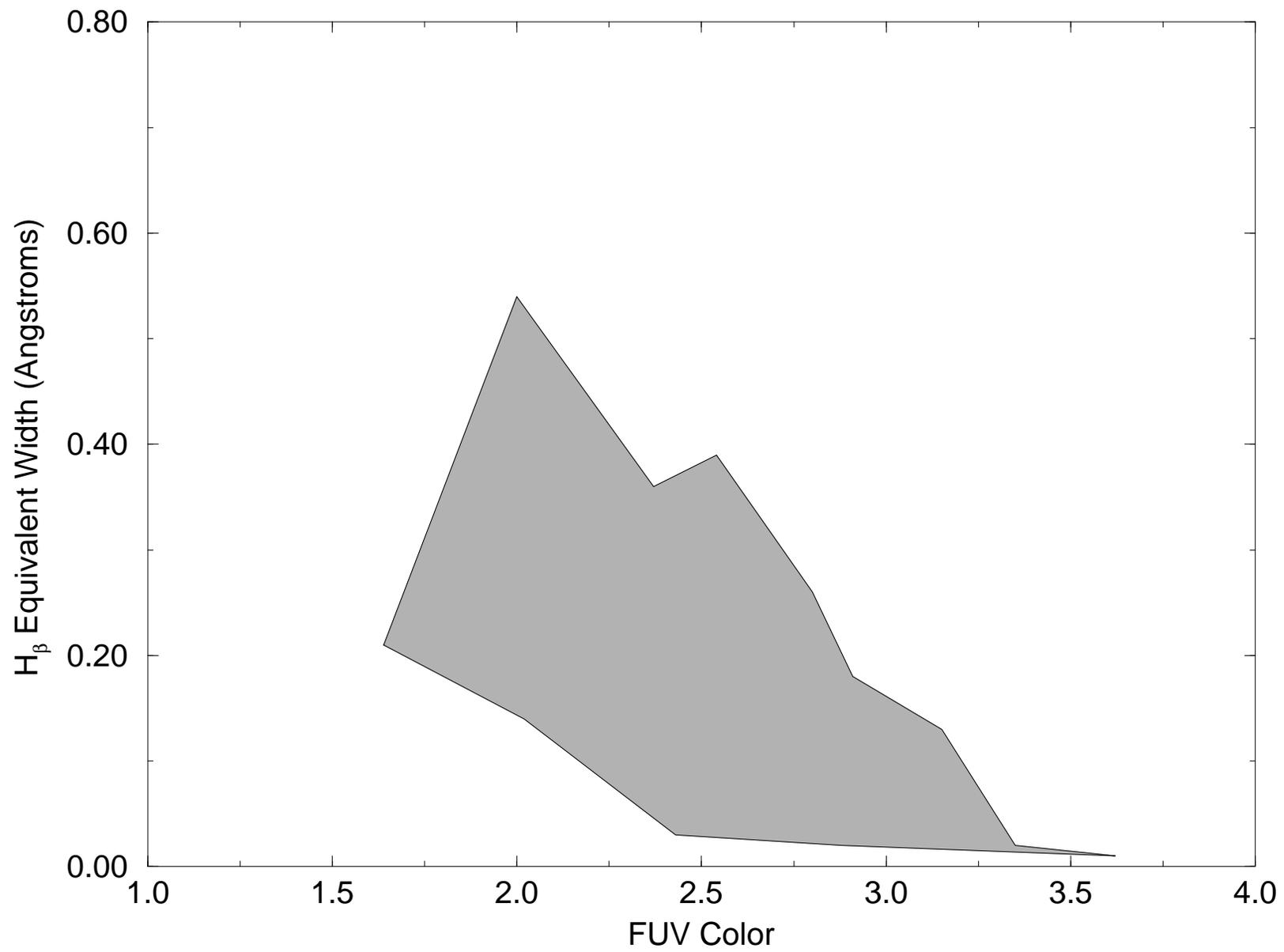
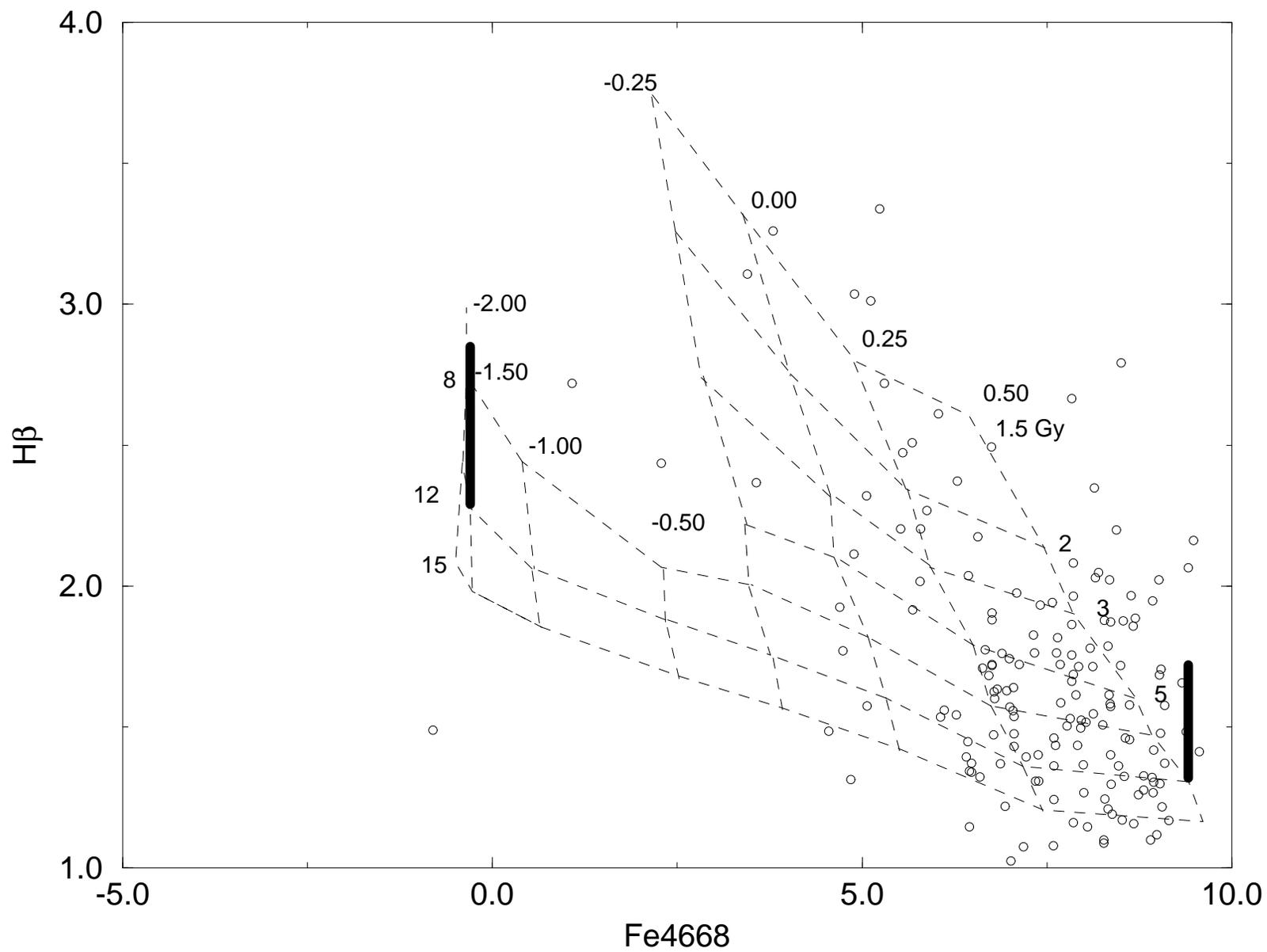
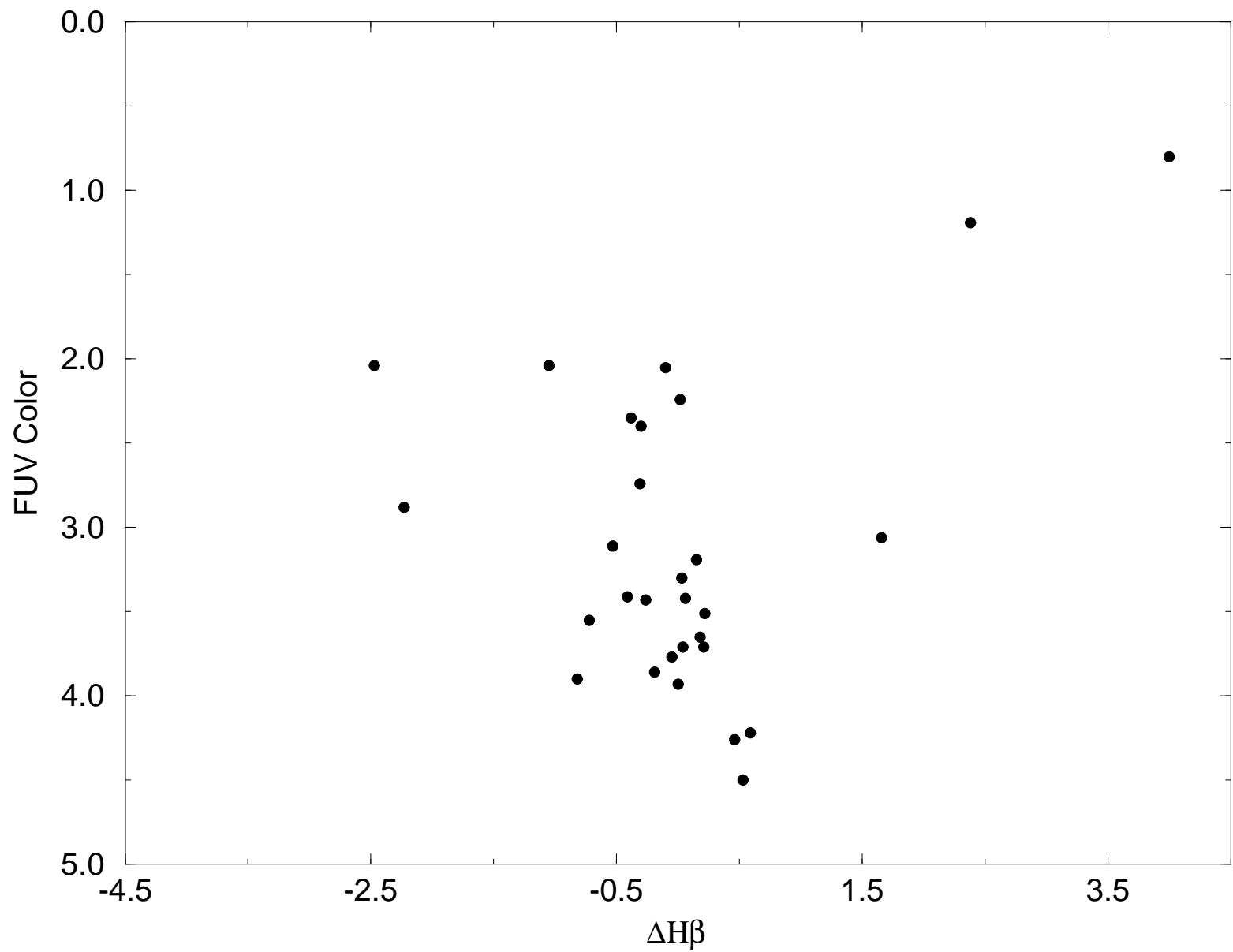
\end{document}